\documentclass[]{interact}

\usepackage{epstopdf}
\usepackage{subfigure}
\usepackage[title]{appendix}

\usepackage{natbib}
\bibpunct[, ]{(}{)}{;}{a}{}{,}
\renewcommand\bibfont{\fontsize{10}{12}\selectfont}

\theoremstyle{plain}

\theoremstyle{definition}

\theoremstyle{remark}

\begin{document}


\title{Tobit Exponential Smoothing, towards an enhanced demand planning in the presence of censored data.}

\author{
\name{D.~J. Pedregal\textsuperscript{a}\thanks{CONTACT D.~J. Pedregal and J.~R. Trapero. Emails: diego.pedregal@uclm.es, juanramon.trapero@uclm.es}; J.~R. Trapero\textsuperscript{b} and E. Holgado\textsuperscript{b}}
\affil{\textsuperscript{a}Industrial Engineering School; \textsuperscript{b}Faculty of Chemical Science and Technology, Universidad de Castilla-La Mancha, Ciudad Real, Spain}}

\maketitle

\begin{abstract}
ExponenTial Smoothing (ETS) is a widely adopted forecasting technique in both research and practical applications. One critical development in ETS was the establishment of a robust statistical foundation based on state space models with a single source of error. However, an important challenge in ETS that remains unsolved is censored data estimation. This issue is critical in supply chain management, in particular, when companies have to deal with stockouts. This work solves that problem by proposing the Tobit ETS, which extends the use of ETS models to handle censored data efficiently. This advancement builds upon the linear models taxonomy and extends it to encompass censored data scenarios. The results show that the Tobit ETS reduces considerably the forecast bias. Real and simulation data are used from the airline and supply chain industries to corroborate the findings.
\end{abstract}


\section{Introduction}
Exponential smoothing (ETS) is one of the techniques most often used in the field of forecasting nowadays, both by practitioners and researchers. These methods have been around since the original work by Brown and Holt in the 1950s. A first critical review dating back to 1985 of ETS can be found in \citep{gardner85}, 20 years later, in 2006, an updated review is done by the same author in \citep{GARDNER2006}. In that latter state of the art, the author indicates that: `the most important theoretical advance is the invention of a complete statistical rationale for exponential smoothing based on a new class of state-space models with a single source of error'. Such an invention was developed by \citep{HYNDMAN2002439}, who provided a novel approach to systematise and automatised exponential smoothing methods by setting such methods in a State Space (SS) framework, that allowed for a rigorous statistical treatment. SS models opened the door to represent ETS methods as stochastic models which permitted to employ likelihood estimation, uncertainty estimation, and model selection techniques among others, see \citep{hyndmanbook2008} and references therein. In addition, the existence of free libraries in specialised software such as R \citep{SvetunkovAdam, hyndman2018forecasting} has also popularised ETS among academics and practitioners alike. 

Typically, the SS for ETS are of the innovations form class, a peculiar setting meaning that there is only one source of noise in the system, in such a way that given an initial state vector, the states are known exactly, with no uncertainty. However, in practice, the initial states are not known, and there are different ways to deal with this matter, as we shall discuss later in Section \ref{sec:censoredLinear}. Often, the single source of noise SS systems are seen as particular cases of multiple source of noise counterparts \citep{SBRANA2020107654}, but the issue is rather more complex, because, under general conditions, single source noise systems have equivalent multiple source representations. This topic is far beyond the scope of this paper, see discussions in  \cite{jerez16}.


One of the most popular applications of ETS in either of its forms has been demand forecasting within supply chain and inventory control problems \citep{PETROPOULOS2022705}. In this case, demand is typically estimated on the basis of recorded sales data. However, the trouble of such measurements is that they do not reflect demand correctly when shortages are produced  \citep{Nahmias1994}. In such situations, real demand is usually higher than sales, and this fact produces generally downward biased forecasts, as well as incorrect uncertainty estimations. Based on this wrong forecasts, safety stocks and reorder points are also wrongly estimated, resulting in lower customer service level, which may cause a spiral-down effect   \citep{Cooper2006}. Nevertheless, the censored demand estimation problem can be applied to other sectors as bike sharing systems \citep{ALBINSKI201859} or hotel revenue management \citep{WEATHERFORD2003401}.


This problem is known as censored estimation in the statistical literature. In fact, James Tobin was the first to analyse censored data in a model relating household income and expenditure \citep{Tobin1958}. The result of this article was the Tobit models, which refer to censored or truncated static regression models with part of the range of the dependent variable constrained. These models can be classified into five categories (Tobit type I - type V) where Tobit type I stands for the model described above. There has been some scarce extensions to dynamic models proposed in the literature long ago in different setups, some of them are not in SS frameworks, e.g., \cite{Zeger1985, Park2007, Wang2018}, while others are in SS frameworks using multiple sources of error \citep{HARVEY2021,Allik16}. However, to the best of authors' knowledge, this topic has not received much attention in the literature. This can be seen in most recent studies on supply chain forecasting that did not even mention this problem though it is acknowledged as a true source of trouble to industry, \citep[see e.g., ][]{BOONE2019170, BOONE2019121, BabaiBoylan2022, Fotios2023, Svetunkov2020, Syntetos2023}. More specifically, there is not any exponential smoothing model developed under a single source of error SS framework which offers a general and elegant solution to the problem of censored data. Note that this open question is also posed by \citep{hyndmanbook2008} in Chapter 18. 

In this paper, we bridge that gap by proposing an exponential smoothing model called Tobit ETS, which is developed under a single source of error SS framework. The main advantage of this development is that we can take advantage of the linear models taxonomy described by \citep{HYNDMAN2002439} and extend it to problems with censored data. Note that, such a taxonomy is mainly described with the acronym ETS, which refers to the different time series components: error, trend and seasonality and can also be considered an abbreviation of \textit{E}xponen\textit{T}ial \textit{S}moothing \citep{hyndmanbook2008}. 

These Tobit ETS models (or TETS) are implemented in the library UComp, with all core functions coded and compiled in C++ and are available for free to the general public in R \citep{UComp}, Python \citep{UcompPy} and Matlab/Octave \citep{UCompMAT}. The library includes automatic identification of TETS models based on the minimisation of information criteria.

The paper is structured as follows: Section \ref{sec:censoredLinear} describes the censored innovations SS model and the Tobit ETS; Section \ref{sec:simulations} shows the Tobit ETS working in practice in some simulations. Finally, Section \ref{sec:conclusion} concludes.

\section{Censored linear Gaussian innovations state space models}
\label{sec:censoredLinear}
Consider the following linear Gaussian innovations state space model

\begin{equation}
\label{eq:sysinnLinear}
\begin{array}{c}
    x_{t} = F x_{t-1} + g \epsilon_{t},\\
    y_t = w x_{t-1} + \epsilon_t,
 \end{array}
\end{equation}

\noindent where $y_t$ is the output time series; $x_t$ is a state vector; $\epsilon_t$ is a scalar perturbation which follows an iid normal distribution with zero mean; and $F$, $w$ and $g$ are system matrices of appropriate dimensions ($g$ is the so called Kalman gain).

To include the Tobit censoring, system \eqref{eq:sysinnLinear} is extended in \eqref{eq:sys} to include the possibility of censorship over the observation variable,  such as:

\begin{equation}
\label{eq:sys}
\begin{array}{l}
y_t^* = w x_{t-1}+ \epsilon_t^*\\
x_{t} = Fx_{t-1}+ g \epsilon_{t}^*\\
y_t= \left\{
    \begin{array}{ll}
       y_{t}^*,  & y_{t}^* \leq Y_{max,t} \\
       Y_{max,t} & y_{t}^* > Y_{max,t}
    \end{array}
    \right.
\end{array}
\end{equation}

\noindent where $y_t^*$ and $y_{t}$ are the uncensored and censored variables, respectively, and $Y_{max,t}$ is the censored level, which is not restricted to be constant and, then, it can be time-varying. The noise in the system is represented by $\epsilon_t^* \sim N(0,\sigma^2)$. 

As happens in system \eqref{eq:sysinnLinear}, if the uncensored $y_t^*$ variable were observed, given a known initial value for the states $x_1$, the rest of states would be known exactly, i.e., 

\begin{equation}
\label{eq:lssrec}
    \begin{array}{c}
    \epsilon_t^*=y_t^*-w x_{t-1}, \\
    x_{t}=F x_{t-1}+ g \epsilon_t^*\\
    \end{array}
\end{equation}

\noindent and the one step ahead forecast would be $\textrm{E} (y_{t}^*|x_{t-1}) = w x_{t-1}$ with variance $\textrm{Var} (y_{t}^*|x_{t-1}) = \sigma^2$, as in (\ref{eq:lssrec}). However, $y_t^*$ is not observed at some time stamps and, then, it is necessary to estimate the moments of the distribution of $y_t^*$ based on the moments of the censored data $y_t$ only.

In summary, to employ the Tobit ETS model we need, first, to initialise the system by computing the states $x_1$ and, second, estimate the moments of the distribution of $y_t^*$. Both problems will be solved in the next subsections.

\subsection{Initialisation}

One feature of ETS models is that the initial state is crucial to estimation, especially in small samples, and it is generally not known. Regarding the way that Tobit ETS can be initialised depends on whether there is censorship in the initialisation sample or not. In case that there is censorship in the initialisation sample, the initial state vector may be treated as a deterministic quantity and estimated along with the rest of parameters by maximum likelihood or by using an information filter as in \citep{hyndmanbook2008}. Nonetheless, in case there is not censorship, we can treat the initial state vector as a multivariate stochastic variable and estimating it by diffuse initialisation, see e.g., \citep{koopman2012} or the augmented Kalman filter (AKF), proposed by \citep{deJong1991}. This procedure is more robust than the deterministic initialisation that in some situations could present numerical inefficiencies  \citep{KohnAnsley1985,deJong1991}. However, to the best of the author's knowledge, such type of diffuse initialisation of the AKF for single source of error state space models has never been developed. Appendix \ref{sec:appA} bridges that gap and provides a general solution to this problem.

\subsection{Estimation}
The main problem is that $y_t^*$ is not observed at some time stamps and then the challenge becomes to estimate the moments of the distribution of $y_t^*$ based on the moments of the censored data $y_t$ only. Appendix \ref{sec:appC} shows that the recursions for the Tobit model become 

\begin{equation}
\begin{array}{rl}
    \textrm{E} (y_{t}|x_{t-1}) = &P_{un,t} (w x_{t-1} - \sigma m_t) +  P_{max,t} Y_{max,t}\\
    \epsilon_t = & y_t - \textrm{E} (y_{t}|x_{t-1}) \\
    \textrm{E}(x_t | Y_{t})= &F x_{t-1} + P_{un,t} / (1 + c_t / P_{un,t} - m_t^2) g \epsilon_t,
\end{array}
\end{equation}

\noindent where $P_{un,t}$ is the probability of observation at time $t$ to be uncensored, $P_{max,t}$ is the probability of an observation to be censored from above, $m_t$ is the inverse Mill ratio at time $t$, and $c_t$ is  another parameter involved in censored distributions (see Appendixes \ref{sec:appB} and \ref{sec:appC}).

Once the states are estimated, forecasts of the uncensored output are estimated taking conditional expectations on system (\ref{eq:sys}), assuming now that the censoring constraint does not apply, i.e.,

\begin{equation}
\begin{array}{rl}
    \hat{x}_{T+1} = &F x_{T} \\
    \hat{y}^*_{T+1} = &w x_{T} \\
    \textrm{Var} (\hat{x}_{T+1})  =& 
    F \textrm{Var} (\hat{x}_{T}) F' + gg' \sigma^2 \\
    \textrm{Var} (\hat{y}_{T+1}^*)  =& 
    w \textrm{Var} (\hat{x}_{T}) w' + \sigma^2 \\
\end{array}
\end{equation}

{Mind that just at the forecast origin $\textrm{Var} (\hat{x}_T) = 0$. Forecasts for more than one step ahead may be calculated by repeating recursively the previous formulas.

Estimation of model parameters is carried out by maximum likelihood including the initial state vector as known fixed values. Taking $\phi$ as the Gaussian standard pdf, the likelihood function for the Tobit model is

\begin{equation}
\label{eq:sys2b}
\begin{array}{c}
L(y_1,\dots,y_n) = \prod_{y_t=Y_{max,t}} P_{max,t} \times \prod_{y_t < Y_{max,t}} \frac{1}{\sigma} \phi \left( \frac{y_t - w x_{t-1}}{\sigma} \right) \\
\end{array}
\end{equation}

\section{Case Studies}
\label{sec:simulations}
To comprehensively evaluate the performance of the Tobit ETS, some simulation studies were conducted using Matlab. For the simulations, synthetic data was generated for cases 1 and 2, while real-world data was employed for cases 3 and 4, in order to provide a thorough assessment of the model's capabilities across diverse scenarios. 

To generate the synthetic data-set utilised in cases 1 and 2, a set of time series comprising 300 observations each is generated. These observations are derived from a Gaussian distribution with a mean of 100 and a standard deviation of 20. The real data-sets used in cases 3 and 4 are: i) the well-known airline passengers time series \citep{boxandjenkins2015}; and ii) a seasonal SKU time series obtained from the M5 series \citep{M5,Makridakis2022}.

Note that, in many real world scenarios, the data that exceed the censored level is not usually registered and, then, it is not possible to quantify the actual error. However, to tackle this issue in the simulations, the observed data (cases 3 and 4) are treated as uncensored data. Subsequently, such data is artificially truncated to obtain the censored data for forecasting purposes.

\subsection{Benchmark models and performance indicators}

In order to assess the efficacy of the Tobit ETS model, its performance is compared against benchmark models. Specifically, Exponential Smoothing is used as the benchmark for cases 1 and 2, whilst Holt-Winters serves as the benchmark for cases 3 and 4.

To evaluate the performance of the models, two commonly used measures for this purpose are used: Root Mean Squared Error (RSME) and Mean Error (ME). RMSE is a widely used metric for evaluating the accuracy (error magnitude) of a prediction or estimation model, such as:
\begin{equation}
\text{RMSE} = \sqrt{\frac{1}{n} \sum_{t=1}^{n}\left[F(t) - y(t)\right]^2}
\end{equation}
\noindent where $y(t)$ is the actual value and $F(t)$ is the forecast.

ME is a measure of systematic error or bias in predictions, i.e.,  

\begin{equation}
    ME = \frac{1}{n} \sum_{t=1}^{n} \left[F(t) - y(t)\right]
\end{equation}

ME can be positive or negative. A positive ME indicates that the predictions tend to be higher than the true values, while a negative ME indicates that the predictions tend to be lower. A ME close to zero indicates minimal bias in the predictions.

In addition to the aforementioned statistical error metrics, regarding real data case IV, lost sales size and excess stock will also be used as performance indicators. Lost sales is the difference between actual demand and its forecast and it occurs when actual demand is higher than its forecast, what implies stock outs. On the other hand, if actual demand is lower than its forecast, a situation of excess stock is happening and it can be calculated as the difference between both values. It is interesting to measure both deviations given that the lost-sales cost usually is higher than excess stock cost.

 \subsection{Case study I - Gaussian distribution with censored data over the mean}
 \label{sec:nordistcensovermean}
In this first simulation, 150 observations of a historical demand are randomly generated with a mean 100, a standard deviation of 20, and a censored level of 120. This simulation corresponds to a Single Exponential Smoothing (SES) model or ETS(A,N,N), meaning that a model is set up with additive error, no trend, and no seasonality \citep{hyndmanbook2008}. In terms of system \eqref{eq:sys}, its SS representation implies $w=1$, $F=1$, $g=\alpha$ (where $\alpha$ is the smoothing constant to be estimated from the data). 



\begin{figure}[htbp]
\centering
    \includegraphics[scale=0.75]{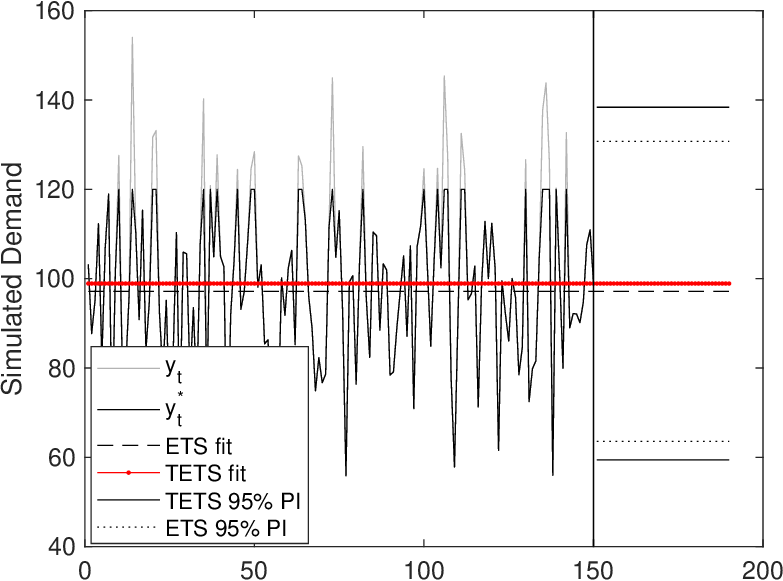}
    \caption{Synthetic demand based on Gaussian distribution. Censored level (120) above the mean of the distribution (100)}
    \label{fig:nordistcensovermean}
\end{figure}

Figure \ref{fig:nordistcensovermean} depicts the censored ($y_t$) and uncensored data ($y_t^*$). It also shows the point forecasts provided by EST and TETS models which are close to 100, although, in the case of ETS, a greater bias down is present with regard to TETS. In addition, prediction intervals at 95\% confidence are also shown at the end of the sample, where the width of intervals provided by ETS is remarkably downward biased. 



\subsection{Case study II - Gaussian distribution with censored data under the mean} \label{sec:nordistcensundermean}
In the second simulation, the experiment is more challenging given that the censored level is set to 90 under the distribution mean (100), what makes that more than 100\% of the data is censored.  Here, the bias introduced by ETS is bigger, however, TETS is able to track the underlying true mean accurately. Again, the 95\% prediction intervals obtained by ETS is totally misleading as can be seen in Figure \ref{fig:nordistcensundermean}.

To evaluate the performance of forecasting models more thoroughly, 10,000 time series are simulated and forecasted 10 periods ahead. Table \ref{tab:SyntheticResultsComparison} shows several mean forecasting error metrics of both the ETS and TETS models across all simulations. In particular, Root Mean Squared Error, forecasting bias (forecasts minus actual values) and the bias of estimated standard deviation of forecasts are included (estimated standard deviation of forecasts minus the simulated one, that is 20).  

Bearing in mind that the simulated mean is 100, the simulations involve different censoring levels, i) no censoring at all, ii) above the mean (120, case I in previous section), iii) just on the mean (100), and iv) below the mean (90, case II in this section). Results clearly show that all forecasts are unbiased when no censoring is present, but all the metrics deteriorate very quickly as the censoring level is more constraining for ETS models in Table \ref{tab:SyntheticResultsComparison}. The metrics are absolutely stable in the case of the Tobit model, regardless of the censoring level.



\begin{table}[h]
    \centering
    \begin{tabular}{lrrrrrr}
        & \multicolumn{2}{c}{\textbf{RMSE}} & \multicolumn{2}{c}{\textbf{Bias}} & \multicolumn{2}{c}{\textbf{SD bias}}\\ \hline
    Censored level    &  TETS   &  ETS    &  TETS    &  ETS     &  TETS    &  ETS   \\ \hline
None & - & 19.6 & -  & -0.0 & -    & 0.0 \\
120 (case I) &  \textbf{19.6} & 19.6 &  \textbf{0.1} &  -1.6 &  \textbf{0.1} &  -2.6 \\
100 &  \textbf{19.6} & 21.0 & \textbf{-0.0} &  -7.9 & \textbf{-0.1} & -8.3 \\
90 (case II) &   \textbf{19.6} & 23.8 & \textbf{-0.1} & -13.9 & \textbf{-0.1} & -11.8 \\
\hline
    \end{tabular}
    \caption{Mean error metrics for different censored levels across 10,000 simulations. Lower error is highlighted in bold}
    \label{tab:SyntheticResultsComparison}
\end{table}


\begin{figure}[htbp]
    \centering
    \includegraphics[scale=0.75]{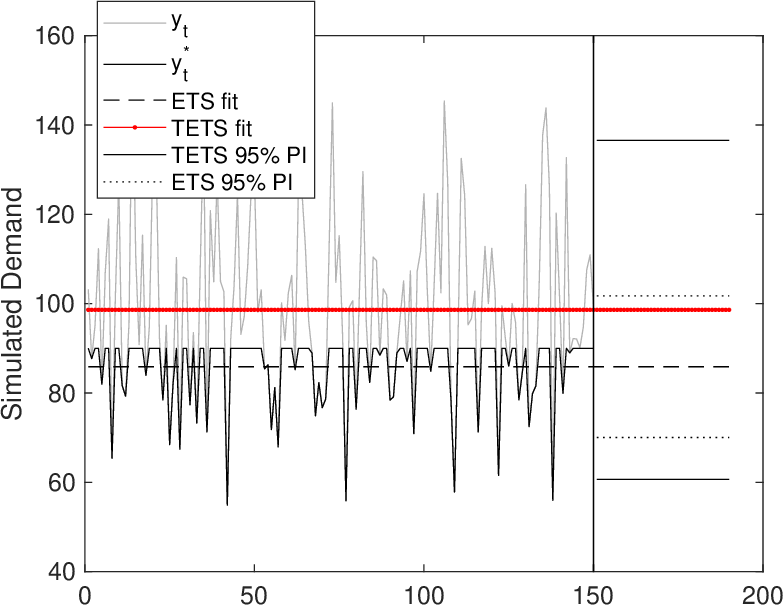}
    \caption{Synthetic demand based on Gaussian distribution. Censored level (90) under the mean of the distribution (100)}
    \label{fig:nordistcensundermean}
\end{figure}




\subsection{Case study III - Airline passengers} \label{sec:airpas}

One of the main benefits of developing TETS is that it takes advantage of the taxonomy defined by \citep{hyndmanbook2008} and, thus, it can be applied to more complex data, which may involve trend or seasonal components. For instance, Figure \ref{fig:Airpassfixcensoredlevel} shows  the log transformation of the well-known air passengers in the US from 1949 to 1960, with an artificial time varying censoring level set at a starting value of 5.6 and increasing linearly from 1956 to the end of the sample. For these data the SES model is no longer valid and other exponential smoothing methods should be chosen. In particular, the Holt-Winters method (ETS(A,A,A)) is considered as a benchmark. In this case, the first 10 years are hold as in-sample data, and the rest of data as forecasted. We can observe in Figure \ref{fig:Airpassfixcensoredlevel} the effect of the censorship over the Holt-Winters forecast, which are biased downwards and totally misleading. On the other hand, the forecast provided by TETS are able to better track the actual values, even in the long run.  


 
\begin{figure}[htbp]
\centering
\includegraphics[scale=0.75]{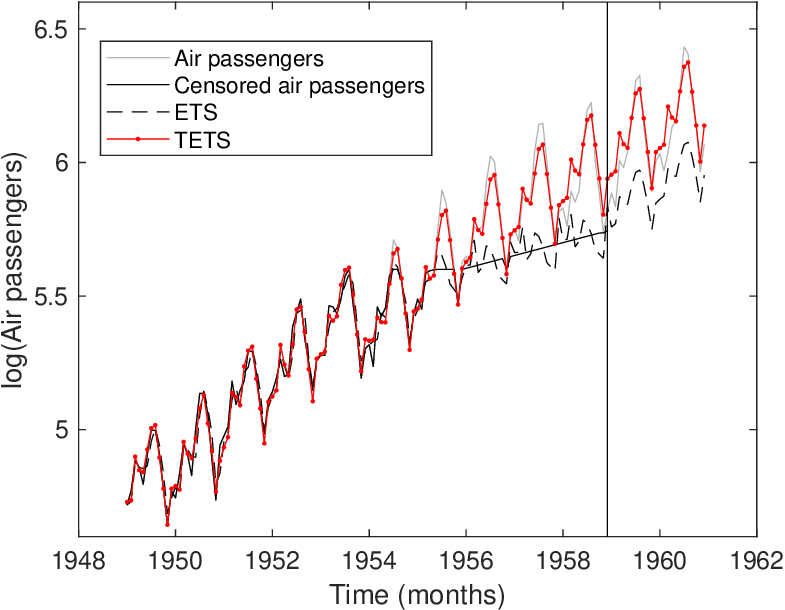}
\caption{Air passengers data in US and time varying censored data and ETS and TETS fits. Forecasts shown from December 1958 onwards.}
\label{fig:Airpassfixcensoredlevel}
\end{figure}




\subsection{Case study IV - Example of M5 dataset} \label{sec:supchaincase}

This case study explores the forecasting of demand under a stock system subject to lost-sales. In this situation, demand that cannot be met is lost and, thus, sales represents an underestimation of demand \citep{SACHS201428}. To simulate this business application, one of the time series from the M5 competition has been used as a true demand \citep{SPILIOTIS2021108237}. In particular, the time series corresponds to the first year of aggregation of all products in level 7, for the state of Wisconsin and department ``\texttt{HOBBIES\_1}''. This time series is in a daily frequency with a typical weekly cycle.

\begin{figure}[htbp]
\centering
\includegraphics[scale=0.75]{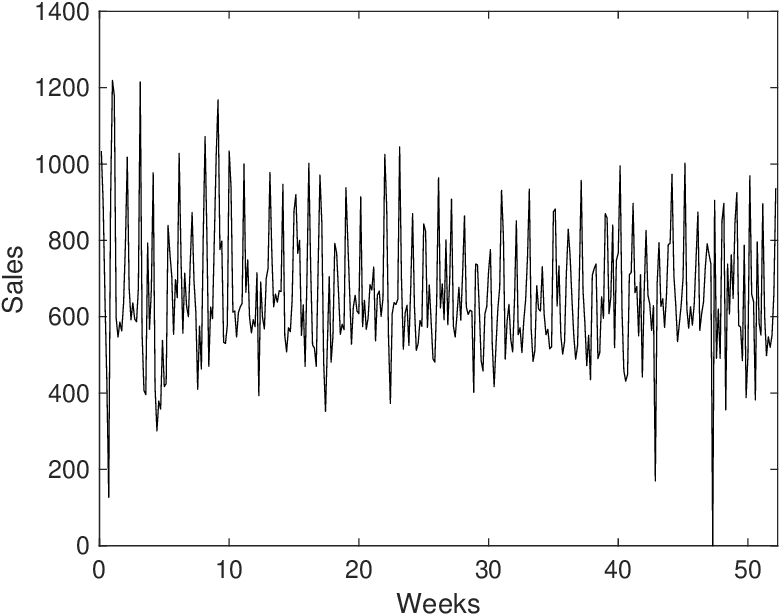}
\caption{Example of seasonal time series from M5 dataset.}
\label{fig:spiralDown}
\end{figure}



Then, such a demand has been censored to obtain the observed sales. Importantly, in this case the censored level is time-varying because it depends on the forecast, which is updated with each new observation. To evaluate in cost terms the implications of the improved forecasts, a newsvendor stock policy is implemented \citep{Axsater2015,Silver2017,JIAN201561}. For this stock policy, the order quantity is defined as the critical fractile ($Y_{max,t+1}$), such as:

\begin{equation}
    Y_{max,t+1}=F_{t+1}+k\sigma_{t+1}
    \label{criticalfrac}
\end{equation}

\noindent $k=\Phi^{-1}(CSL)$ is the safety factor based on the cycle service level (CSL) and $\Phi(\cdot)$ is the standard normal cumulative distribution function. Note that, the safety factor is defined by a service level metric. For the sake of simplicity, the CSL \citep{Silver2017} is chosen, although other metrics as the fill-rate can be implemented. The forecasting models provides both, the mean demand forecast ($F_{t+1}$) and its std deviation forecast ($\sigma_{t+1}$). Expression \eqref{criticalfrac} shows the relationship between the censored level/critical fractile and the forecasts and, more importantly, how the hypothesis of a constant censored level is not realistic in this application. Additionally, that expression also shows that, although the CSL is constant, the censored level can be time-varying.

Table \ref{tab:RealDataResultsComparisonmetrics} presents forecasted metrics from various experiments using different CSL values ranging from 70\% to 99\%. Notably, the root mean squared error (RMSE), which measures forecasting precision, clearly indicates that TETS is consistently more accurate regardless of the CSL. Interestingly, the precision of both models tends to converge as the CSL increases. This is an expected outcome, since the historical demand estimated by both models tend to be closer at the cost of big excess inventory that allows for the observation of the true demand. Similarly, the bias of the forecasts (columns 4 and 5 in Table \ref{tab:RealDataResultsComparisonmetrics}) exhibits analogous behavior, with ETS consistently performing worse but tending to converge towards TETS as the CSL increases.

\begin{table}[h]
    \centering
    \begin{tabular}{rrrrr}
    \hline
CSL  &  RMSE ETS  & RMSE TETS  & Bias ETS & Bias TETS \\
\hline
70\% &     -  &   122.6   &      -  &   -1.3  \\
80\% & 139.2 &    121.9  &   -42.3 &    -6.6  \\
90\% & 124.9  &   121.8  &   -16.0  &   -8.3  \\
95\% & 123.0  &   121.2  &   -11.8 &    -9.3  \\
99\% & 121.2  &   120.3  &    -9.9  &   -9.1  \\
    \hline
    \end{tabular}
    \caption{Performance metrics for the case study IV for different objective cycle service levels.}
    \label{tab:RealDataResultsComparisonmetrics}
\end{table}
The case of a CSL of 70\% is an exception that deserves special attention. The reason is that a spiral down happens and the ETS forecasts lose all meaning, as shown in Figure \ref{fig:spiralDown}. This figure shows that demand built on the basis of one-day-ahead forecasts for each model (ETS and TETS) perfectly matches the true demand during the initial 5 weeks, because for all those days excess stock happens systematically and hence the true demand is observed. However, as soon as a stockout happens, the estimated censored demands start to drift away from true demand and from each other. It is remarkable that both models forecast demand poorly during the fifth week, but ETS never recovers from that bad forecast. The reason is that the bad forecast led to lost sales that produced an insufficient inventory level for the next day, which again led to underforecast demand, and so on. It is remarkable that the TETS model does not actually have this risk, because the forecasts are unbiased.

\begin{figure}[htbp]
\centering
\includegraphics[scale=0.75]{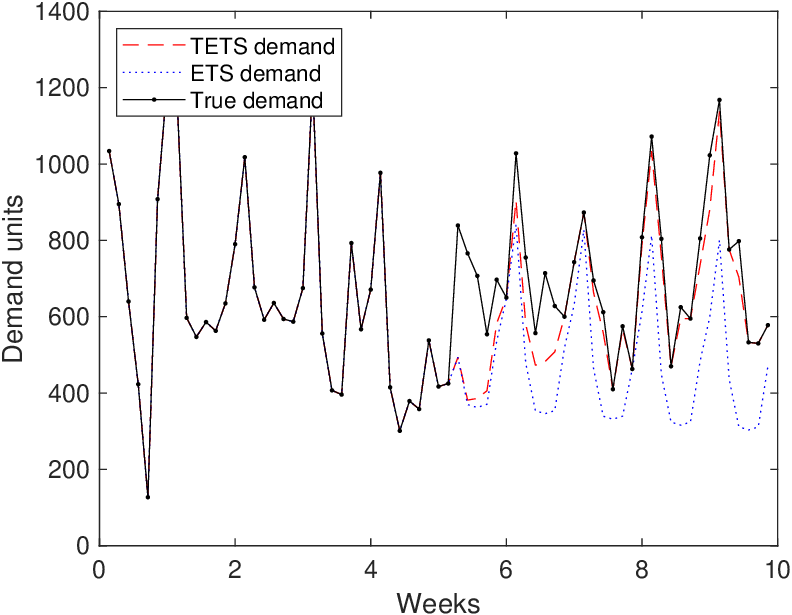}
\caption{TETS, ETS and true demand for M5 data.}
\label{fig:spiralDown}
\end{figure}

Finally, and most importantly, Table~\ref{tab:RealDataResultsComparisonKPI} presents the lost sales, excess stock, and achieved CSL for each proposed target CSL (first column). The TETS model generates fewer lost sales at the expense of more excess stock, but it also achieves a higher CSL. As the target CSL increases, both models naturally converge towards similar performance. However, the standard ETS model carries a risk of entering spiral down loops, which, in this specific dataset with its relatively stable demand, occur at a surprisingly low target CSL of 70\%.

\begin{table}[h]
    \centering
    \begin{tabular}{rrrrrrr}
    \hline
CSL  &  LS ETS &  LS TETS &  EXS ETS & EXS TETS &  CSL ETS & CSL TETS \\
    \hline
70\%  &    -  &    7,301  & - &    35,760   &    - &  75.0\% \\
80\% & 10,817 & 4,349  & 31,148  & 46,842 &  69.4\% &  85.1\% \\
90\% & 2,605  & 1,870 &  59,805 &  66,732 &  90.2\%  & 92.9\% \\
95\% & 1,059  & 922  &  80,731 &  83,953 &  96.2\%  & 96.7\% \\
99\% & 216  & 197  &  118,019 &  119,843 & 98.9\% & 98.9\% \\
    \hline
    \end{tabular}
    \caption{Lost sales (LS) excess stock (EXS) and cycle service level (CSL) for both ETS and TETS models.}
    \label{tab:RealDataResultsComparisonKPI}
\end{table}

\section{Conclusions}
\label{sec:conclusion}
Exponential smoothing methods have been widely used since the 50s. However, a crucial moment in the development of ETS was the introduction of a statistical rationale based on state space models that enabled rigorous statistical treatments. Such a development and its implementation in free software toolboxes has allowed to make ETS models a preferred forecasting tool in many applications. Nevertheless, its use when facing a censored signal remained as an open question.

This work has developed the Tobit ETS under a single source of error state space framework. This model is capable of dealing with a either constant or time-varying censorship. In fact, thanks to this contribution, all the previous work made in the ETS literature can be extended to cope with censored signals. 

In addition, this study have compared the performance of the Tobit ETS against other well-known benchmarks as the Single Exponential Smoothing and Holt-Winters methods. Simulated and actual data have demonstrated that the Tobit ETS outperformed all the considered benchmarks in all the cases analyzed. 

Further research should address the utilization of the Tobit ETS to empirical case studies from different business problems. One potential application is related to supply chain forecasting demand for companies that face lost-sales in case of stockouts.





\section*{Funding}
This work was supported by the Vicerrectorado de Investigación y Política Científica from UCLM through the research group fund program (PREDILAB; DOCM 2023-4428 [2022-GRIN-34368]).








\bibliographystyle{tfcad}
\bibliography{sample}

\begin{appendices}
\section{Linear innovations state space models}
\label{sec:appA}

Consider the following innovations state space model

\begin{equation}
\label{eq:sysinnLinearA}
\begin{array}{c}
    x_{t} = F x_{t-1} + g \epsilon_{t},\\
    y_t = w x_{t-1} + \epsilon_t,
 \end{array}
\end{equation}

\noindent with $\epsilon_t \sim N(0,\sigma^2)$. Assuming $Y_{t}$ represents all the information available up to and including observation $t$, the innovation ($v_t$) is just the perturbation in the equations, i.e.,

\begin{equation}
    v_t=y_t-\textrm{E}(y_t|Y_{t-1})=w x_{t-1} + \epsilon_t-\textrm{E}(w x_{t-1}+\epsilon_t|Y_{t-1})= \epsilon_t.
\end{equation}

The Kalman filter applied to this system colapses to the following recursions, that may be derived directly from the system

\begin{equation}
\label{eq:KFinnA}
    \begin{array}{c}
    \epsilon_t=y_t-w x_{t-1}, \\
    x_{t}=F x_{t-1}+ g \epsilon_t
    \end{array}
\end{equation}

The main point in the innovations systems is that, assuming a known initial state, the subsequent states are observed with no uncertainty. However, still an initial state is needed and, therefore, the estimated states are estimated conditional on such initial state.



One way out of this difficulty is using diffuse initialization. A general treatment of diffuse initialization is the following, where $q$ is the number of non-stationary elements agglutinated in $\delta$ vector,

\begin{equation}
\begin{array}{cc}
    x_1 = x+A\delta+R\epsilon_0, &  \epsilon_0 \sim N(0,Q_0) \\
    \delta \sim N(0,\kappa I_q) & \kappa \to \infty
\end{array}
\end{equation}

$A$ is a selection matrix that is simply relating the non-stationary states ($\delta$) to the general initial state vector $x_1$, and $R$ is playing a similar role, but with the stationary states ($\epsilon_0$). A convenient re-arrangement of columns of matrices $A$ and $R$ would produce an identity matrix. $Q_0$ is the variance of stationary states.

The development of the AKF for SS innovations form involves the standard KF recursions with zero initial states (i.e., equations (\ref{eq:KFinnA})) and two additional recursions to deal with non-stationary states \citep{deJong1991}. Such algorithm in the single source of error case reduces to

\begin{equation}
\label{eq:AKFinnAA}
    \begin{array}{c}
    \hat{\epsilon}_t^-=y_t-w x_{t-1}^-, \\
    x_{t}^-=F x_{t-1}^-+ g \epsilon_t^-\\
    V_t = -w A_{t-1} \\
    A_{t} = F A_{t-1} + g V_t
    \end{array}
\end{equation}

\noindent with $A_{1}=-A$ and $\epsilon_t^- \sim N(0,\sigma^{2-})$. All the minus superscripts are included to signal that these quantities are conditional on the arbitrary initialization. Defining

\begin{equation}
\label{eq:AKFllikInnA}
    \begin{array}{rl}
    s_t= &\sum_{j=1}^t V_j\epsilon_j^- / \sigma^{2-},\\
    S_t= &\sum_{j=1}^t V_j V_j' / \sigma^{2-},
    \end{array}
\end{equation}

\noindent the filtered states are finally estimated as

\begin{equation}
\label{eq:AKFllikInnAA}
    \begin{array}{c}
    \hat{x}_t= \hat{x}_t^- - A_t S_t^{-1}s_t.
    \end{array}
\end{equation}

The likelihood may be written as

$$L(y_1,y_2,\dots,y_n)=-0.5 [ (n-k)(\ln \hat{\sigma}^2 + 1) + \sum_{t=1}^n\ln \sigma^{2-} + \ln |S_n| ] $$

\noindent with $\hat{\sigma}^2 = 1/(n-k)\left[ \sum_{t=1}^n \epsilon_t^{2-} / \sigma^{2-}-s_n'S_n^{-1}s_n \right]$.

Therefore, the initial state vector estimated conditional on the whole sample is $\hat{x}_1=S_n^{-1}s_n$ \citep[see][]{deJong1991}.

\section{Truncated and censored Gaussian random variables}
\label{sec:appB}

Consider a Gaussian variable $x \sim N(\mu, \sigma)$. If a maximum truncation limit is set at $a$, then, the censored pdf is 

\begin{equation}
f(x_C)= \left\{
    \begin{array}{ll}
       f(x) = \frac{1}{\sigma} \phi((x - \mu) / \sigma),  & x \leq a \\
       1-\Phi(\alpha), & x > a
    \end{array}
    \right.,
\end{equation}

\noindent where $\phi$ and $\Phi$ stand for the standard normal pdf and cdf, respectively, and $\alpha=(a-\mu)/\sigma$. The truncated pdf is

\begin{equation}
x_T= \left\{
    \begin{array}{ll}
       f(x) / \Phi(\alpha) = \frac{1}{\sigma} \frac{\phi((x - \mu) / \sigma)}{\Phi(\alpha)},  & x \leq a \\
       0, & x > a
    \end{array}
    \right..
\end{equation}

Now

\begin{equation}
    \textrm{E} (x_T) = 1 / \Phi(\alpha) \int_{-\infty}^{a} x f(x) dx = \mu - \sigma \phi(\alpha) / \Phi(\alpha)
\end{equation}

\begin{equation}
    \textrm{Var} (x_T) = \sigma^2 [1 - \alpha \phi(\alpha) / \Phi(\alpha) - (\phi(\alpha) / \Phi(\alpha))^2]
\end{equation}

\begin{equation}
\begin{array}{rl}
   \textrm{E} (x_C) = & P(x>a) a + P(x \leq a) \textrm{E} (x|x \leq a) = \\
     = &  [1-\Phi(a)] a + \Phi(a) \textrm{E} (x_T)
\end{array}
\end{equation}

\begin{equation}
   \textrm{Var} (x_C) = \textrm{Var} (x_T)
\end{equation}

\section{Censored linear Gaussian innovations state space models}
\label{sec:appC}
The linear Gaussian innovations state space model with Tobit censoring from above, where $y_t^*$ is the uncensored variable and $y_t$ is its censored counterpart is

\begin{equation}
\label{eq:sysA}
\begin{array}{l}
y_t^* = w x_{t-1}+ \epsilon_t^*\\
x_{t} = Fx_{t-1}+ g \epsilon_{t}^*\\
y_t= \left\{
    \begin{array}{ll}
       y_{t}^*,  & y_{t}^* \leq Y_{max,t} \\
       Y_{max,t} & y_{t}^* > Y_{max,t}
    \end{array}
    \right.
\end{array}
\end{equation}

\noindent and $\epsilon_t^* \sim N(0,\sigma^2)$. If the uncensored $y_t^*$ variable were observed, given a known initial value for the states $x_1$ the rest of states would be known exactly and the one step ahead forecast would be $\textrm{E} (y_{t}^*|x_{t-1}) = w x_{t-1}$ with variance $\textrm{Var} (y_{t}^*|x_{t-1}) = \sigma^2$. The limiting constraints would not apply and forecasting the output would consist of applying the following recursions

\begin{equation}
\label{eq:recurs}
\begin{array}{rl}
    \textrm{E} (y_{t}^*|x_{t-1}) = & w x_{t-1} \\
    \epsilon_t^*= & y_t^* - \textrm{E} (y_{t}^*|x_{t-1}) \\
    \textrm{E} (x_t|Y_t^*) = x_t = & F x_{t-1} + g \epsilon_t^*.
\end{array}
\end{equation}

The problem is that $y_t^*$ is not observed at some time stamps and the analysis should be based on the observed data $y_t$ only. Define $P_{un,t}$ and $P_{max,t}$ as the probabilities of observation at time $t$ to be uncensored or censored from above, respectively. Based on the moments of censored random variables (see \ref{sec:appB}) the two first expressions in (\ref{eq:recurs}) convert to

\begin{equation}
\begin{array}{rl}
    \textrm{E} (y_{t}|x_{t-1}) = &P_{un,t} (w x_{t-1} - \sigma m_t) + P_{max,t} Y_{max,t},\\
    \epsilon_t = & y_t - \textrm{E} (y_{t}|x_{t-1}), \\
\end{array}
\end{equation}

\noindent with definitions:

$$ P_{max,t} = 1 - \Phi\left(\frac{Y_{max,t}-w x_{t-1}}{\sigma}\right),$$

$$ P_{un,t}= 1 - P_{max,t},$$

\noindent and $\Phi$ standing for the standard Gaussian cdf. 

Note that $\epsilon_t$ is the censored innovation. The equivalent expression for last equation in (\ref{eq:recurs}), namely for $\textrm{E}(x_t|Y_t)$, requires a bit more of elaboration. Using the regression lemma

\begin{equation}
\label{eq:ExtYt}
\begin{array}{rl}
   \textrm{E}(x_t | Y_{t}) = &\textrm{E}(x_t | Y_{t-1}, \epsilon_t) = \\
   = & \textrm{E}(x_t | Y_{t-1}) + \textrm{Cov}(x_t,\epsilon_t | Y_{t-1})[\textrm{Var}(\epsilon_t| Y_{t-1})]^{-1}\epsilon_t.
\end{array}
\end{equation}

Note that

\begin{equation}
\label{eq:Covxtet}
\begin{array}{rl}
  \textrm{Cov}(x_t,\epsilon_t | Y_{t-1}) = & \textrm{E} [(x_t - \textrm{E} (x_t | Y_{t-1})) \epsilon_t]\\
  = & \textrm{E}[(F x_{t-1} + g \epsilon_t^* - F x_{t-1}) \epsilon_t] \\
  = & \textrm{E} (g \epsilon_t^* \epsilon_t) \\
  = & P_{un,t} g \sigma^2.
\end{array}
\end{equation}

In the uncensored region $\epsilon_t^* = \epsilon_t$. This happens with probability $P_{un, t}$. In the censored region $\epsilon_t$ becomes a constant random variable, meaning that any covariance is zero. Based on \ref{sec:appB}, we have

\begin{equation}
\label{eq:varetYt}
\begin{array}{c}
  \textrm{Var} (\epsilon_t|Y_{t-1})=\textrm{Var} (y_t|Y_{t-1})=\sigma^2 (1 + c_t / P_{un,t} - m_t^2),
\end{array}
\end{equation}

\noindent with (see \cite{Allik16})

$$ m_t=\phi\left(\frac{Y_{max,t}-w x_{t-1}}{\sigma}\right) / P_{un,t}$$

$$ c_t= - \frac{Y_{max,t}-w x_{t-1}}{\sigma}\phi\left(\frac{Y_{max,t}-w x_{t-1}}{\sigma}\right).$$

Putting together equations (\ref{eq:ExtYt}), (\ref{eq:Covxtet}) and (\ref{eq:varetYt}) we have

\begin{equation}
    \textrm{E}(x_t | Y_{t})=F x_{t-1} + P_{un,t} / (1 + c_t / P_{un,t} - m_t^2) g \epsilon_t.
\end{equation}

For convenience we repeat here the whole algorithm with Tobit censoring

\begin{equation}
\begin{array}{rl}
    \textrm{E} (y_{t}|x_{t-1}) = &P_{un,t} (w x_{t-1} - \sigma m_t) +  P_{max,t} Y_{max,t}\\
    \epsilon_t = & y_t - \textrm{E} (y_{t}|x_{t-1}) \\
    \textrm{E}(x_t | Y_{t})= &F x_{t-1} + P_{un,t} / (1 + c_t / P_{un,t} - m_t^2) g \epsilon_t
\end{array}
\end{equation}

\noindent with all given definitions.

The Tobit Kalman filter for this model is developed under the assumption that the state prediction is a sufficiently accurate estimation of censoring from above probability \citep{AllikThesis14}. This poses a big problem for diffuse initialization of any kind either \cite{deJong1991} or \cite{koopman2012}, because they rely on arbitrary initial states, thence, providing unrealistic estimations of censoring probabilities at the beginning of the sample. As we have mentioned in subsection 2.1, in case that censored data is present in the initialization sample, the initial state vector is treated as deterministic quantities and estimated along with the rest of model parameters by maximum likelihood.


\end{appendices}

\end{document}